\documentclass[a4paper,11pt]{article}
\pdfoutput=1 

\usepackage{jheppub} 

\usepackage[T1]{fontenc} 

\newcommand{\be}{\begin{equation}}
\newcommand{\ee}{\end{equation}}
\newcommand{\bea}{\begin{eqnarray}}
\newcommand{\eea}{\end{eqnarray}}

\newcommand{\R}{r_{\rm KK}}
\newcommand{\RSS}{R_{\rm D4}}
\newcommand{\MKK}{M_{\rm KK}}
\newcommand{\UKK}{U_{\rm KK}}

\newcommand{\rKK}{r_{\rm KK}}
\newcommand{\gYM}{g_{\rm YM}}

\newcommand{\FYM}{F_{\rm YM}}

\newcommand{\gs}{g_{s}}

\newcommand{\Tr}{{\rm Tr}\,}
\newcommand{\1}{\zeta_1}
\newcommand{\2}{\zeta_2}
\newcommand{\9}{\bar{\zeta}_1}

\newcommand{\T}{\mathbb{P}}

\title{
Witten-Veneziano mechanism and pseudoscalar glueball-meson mixing in holographic QCD}

\author{Josef Leutgeb}
\author{and Anton Rebhan}
\affiliation{Institut f\"{u}r Theoretische Physik, Technische Universit\"{a}t Wien,\\
Wiedner Hauptstr.~8-10, A-1040 Vienna, Austria}

\emailAdd{josef.leutgeb@tuwien.ac.at}
\emailAdd{anton.rebhan@tuwien.ac.at}

\abstract{We revisit the U(1)$_A$ anomaly in the holographic model of low-energy QCD
by Witten, Sakai, and Sugimoto, presenting a new and direct derivation of the Witten-Veneziano
mechanism for generating the mass of the $\eta'$ through an anomalous mixing
of the Ramond-Ramond $C_1$ field with the singlet component of the pseudoscalar mesons.
The latter turns out to have a kinetic mixing with the normalizable modes of the $C_1$ field
representing pseudoscalar glueballs, 
yielding additional vertices for their production and their decay
that dominate over those of the unmixed case considered previously in the Witten-Sakai-Sugimoto model.
The leading channel is predicted to be decay into two vector mesons, followed in importance
by decay into three pseudoscalar mesons. The issue of production of pseudoscalar glueballs in radiative $J/\psi$ decays
and in double diffractive processes is also discussed briefly.
}

\begin{document} 
\maketitle
\flushbottom

\global\long\def\xtau{\tau}

\global\long\def\xtaui{\tau}

\global\long\def\d{d} 

\global\long\def\Ukk{U_{\text{KK}}}

\global\long\def\rkk{r_{\text{KK}}}

\global\long\def\kk{\text{KK}}

\global\long\def\intxtau{\text{\ensuremath{\int d^{4}}x \,d\ensuremath{\xtau}}}

\global\long\def\intx{\int d^{4}x}

\global\long\def\gs{\Phi}

\global\long\def\kzt{\kappa_{10}^{2}}

\section{Introduction}

In QCD, the U(1)$_A$ part of the tree-level flavor symmetry $\text{U}(N_f)_L\times\text{U}(N_f)_R$ is
broken by the axial anomaly, leading to only $N_f^2-1$ pseudoscalar Goldstone bosons 
in the spontaneous breaking of $\text{U}(1)_V\times \text{SU}(N_f)_L\times\text{SU}(N_f)_R\to \text{U}(N_f)_V$
with masses determined by the
Gell-Mann--Oakes--Renner relation, while one isoscalar pseudoscalar
boson, the $\eta'$, is found to be too heavy to be a Goldstone boson \cite{Weinberg:1975ui}.
Its mass is determined by nonperturbative effects involving the axial anomaly, and it was
suggested by 't Hooft \cite{tHooft:1976rip,tHooft:1986ooh} that these are due to instantons.
However, in the limit of large number of colors $N_c$ the effects of a dilute gas of instantons
is exponentially suppressed. A different mechanism was proposed by Veneziano \cite{Veneziano:1979ec} and Witten \cite{Witten:1979vv}, who
showed that the nontrivial $\theta$-dependence of large-$N_c$ pure Yang-Mills theory implies
a mass term for the singlet $\eta_0$ according to
$
m_0^2={2N_f}\chi_g/{f_\pi^2},
$
where $\chi_g$ is the topological susceptibility of pure Yang-Mills theory.

In \cite{Armoni:2004dc,Barbon:2004dq} it was shown that this mechanism is indeed realized in the AdS/CFT framework.
In the Witten model \cite{Witten:1998zw} of low-energy QCD, which is based on
a supersymmetry-breaking circle compactification of $N_c$ D4 branes in type-IIA supergravity,
this was initially discussed in a model with flavor D6 branes \cite{Barbon:2004dq}
and subsequently also in the chiral
Witten-Sakai-Sugimoto (WSS) model \cite{Sakai:2004cn,Sakai:2005yt}, which is based on D8 branes
and which realizes a fully nonabelian chiral symmetry breaking.

A more detailed discussion of the Witten-Veneziano mechanism in the WSS model was
recently given by Bartolini et al. in \cite{Bartolini:2016dbk}. In the present paper
we present an alternative, more direct derivation, which moreover allows us to reanalyse
the possible mixing of the singlet $\eta_0$ with the pseudoscalar glueball $\tilde G$.
In the bottom-up V-QCD model of ref.~\cite{Jarvinen:2011qe}, where the Veneziano limit of $N_c\to\infty$
with $N_f/N_c$ fixed is employed, such mixing in the context of the Witten-Veneziano relation has
been discussed in \cite{Arean:2016hcs}.
In the WSS model, one of us with F.\ Br\"unner has found \cite{Brunner:2016ygk} a vanishing
mass mixing of $\eta_0$ and $\tilde G$ which led to the conclusion of a very narrow
pseudoscalar state, because to leading order in the WSS model only vertices involving jointly
the scalar glueball and $\eta_0$ appeared to be present. Our new derivation reveals
the presence of a kinetic (derivative) mixing which leads to additional decay modes of pseudoscalar glueballs that are dominating 
those considered in Ref.~\cite{Brunner:2016ygk}
or in the phenomenological model of Ref.~\cite{Eshraim:2012jv}.

The importance of gluonic contributions for the physics of $\eta'$ mesons is
already evidenced by the Witten-Veneziano mechanism. It has also been emphasized
by the analysis of ref.\ \cite{Shore:1991np}, who have however left open the question whether
this involves a coupling to physical pseudoscalar glueball excitations (see also \cite{Ball:1995zv,Harland-Lang:2013ncy,Bass:2018xmz}).
Possible pseudoscalar glueball-meson mixing scenarios have been discussed
widely in the literature \cite{Rosenzweig:1981cu,Rosenzweig:1982cb,Escribano:2008rq,Mathieu:2009sg,Ambrosino:2009sc,Ke:2011fj}, but exclusively in the form of mass mixings.
The mixing scenario that we obtain in the WSS model leads to a dominance
of the couplings that are induced by the Chern-Simons term,
in particular to two vector mesons. Inclusion of quark masses leads to
additional couplings such as to three pseudoscalar mesons.

Assuming that the couplings obtained in the WSS model are more reliable than the results
for the glueball masses, which for tensor and pseudoscalar glueballs are much lower
than those predicted by lattice QCD, we consider a range of pseudoscalar glueball
masses, with the result that a pseudoscalar glueball of around 2.6 GeV
as indicated by (quenched) lattice QCD is found to be a rather broad resonance
instead of a narrow state, which may hinder its experimental identification. 
On the other hand, its relatively strong coupling to vector mesons
should enhance its production cross section in central exclusive production
as well as in radiative $J/\psi$ decays.

In sec.\ \ref{sec2} we review the WSS model, in particular the role of the Ramond-Ramond $C_1$ 
which determines the $\theta$ parameter of the dual theory, and how the U(1)$_A$ anomaly
and the Witten-Veneziano mechanism is realized. In sec.\ \ref{sec:mixing} we derive the mixing
of $\eta_0$ and the pseudoscalar glueball modes contained in the normalizable $C_1$ modes,
and we work out its consequences for glueball-meson vertices and decay patterns of the pseudoscalar
glueball in sec.\ \ref{sec:decay}, followed by a discussion and an outlook in sec.\ \ref{sec:conclusion}.

\section{Witten-Sakai-Sugimoto model and Witten-Veneziano mechanism}\label{sec2}

\subsection{The Witten model of low-energy QCD}

The WSS model is based on the Witten model of low-energy QCD \cite{Witten:1998zw} provided by the near-horizon geometry of
a large number ($N_c$) of coincident D4 branes in type-IIA superstring theory wrapped on a circle of circumference $R_4=2\pi\MKK^{-1}$ with anti-periodic boundary
conditions for fermions. Since gauginos are massive at tree-level and adjoint scalars acquire masses through loops, 
the dual theory at energies much smaller than $\MKK$ is pure non-supersymmetric Yang-Mills theory
at large 't Hooft coupling $\lambda=N_c\gYM^2$. In the supergravity approximation, where the Kaluza-Klein mass scale $\MKK$ is kept finite,
the background can be obtained from the dimensional reduction of an 11-dimensional doubly Wick-rotated black-hole geometry
in AdS$_7\times S^4$,
\be
ds^2_{11}=\frac{r^2}{L^2}\left[ f(r)dx_4^2+\eta_{\mu\nu}dx^\mu dx^\nu+dx_{11}^2 \right]+\frac{L^2}{r^2}\frac{dr^2}{f(r)}+\frac{L^2}{4}d\Omega_4^2,
\quad f(r)=1-\frac{\rKK^6}{r^6},
\ee
where $\mu,\nu=0,\ldots,3$, $\eta_{\mu\nu}=\text{diag}(-1,1,1,1)$,
together with a nonzero 4-form field strength $F_4=3(L/2)^3\omega_4$, where $\omega_4$ is the volume form of a unit 4-sphere with volume $V_4=8\pi^2/3$.
A regular Euclidean black hole is produced by identifying $\tau\equiv x_4 \simeq x_4+2\pi R_4$ with $R_4=\MKK^{-1}=L^2/(3\rKK)$.

Dimensional reduction from 11-dimensional supergravity \cite{Becker:2007zj} with $\kappa_{11}^2=(2\pi l_P)^9/(4\pi)$ and $l_P=g_s^{1/3}l_s$
through $x_{11}\simeq x_{11}+2\pi R_{11}$, with $R_{11}=g_s l_s$, and
\be\label{ds210from11}
ds^2_{11}=G_{\hat M \hat N}dx^{\hat M} dx^{\hat N}=e^{-2\Phi/3}g_{MN}dx^M dx^N+e^{4\Phi/3}(dx^{11}+A_M dx^M)^2,
\ee
where $e^\Phi=(r/L)^{3/2}$ and hatted (unhatted) indices refer to 11 (10) dimensions,
leads to type-IIA supergravity 
with string-frame action
\begin{align}
S= & S_\mathrm{NS}+S_\mathrm{R}+S_\mathrm{CS},
\end{align}
where
\begin{align}
S_\mathrm{NS}= & \frac{1}{2\kappa_{10}^{2}}\int\d^{10}x\sqrt{-g}e^{-2\Phi}\left(R+4\,\partial_{M}\Phi\partial^{M}\Phi-\frac{1}{2}\left|dB_2\right|^{2}\right),
\nonumber \\
S_\mathrm{R}= & -\frac{1}{4\kappa_{10}^{2}}\int\d^{10}x\sqrt{-g}\left(\left|F_{2}\right|^{2}+\left|F_{4}\right|^{2}\right),
\quad |F_p|^2=\frac1{p!}F_{M\ldots}F^{M\ldots},\nonumber \\
S_\mathrm{CS}= & -\frac{1}{4\kappa_{10}^{2}}\int B_{2}\wedge F_{4}\wedge F_{4}.
\end{align}
In order to have a standard form of D-brane actions with
prefactors $\mu_p=(2\pi)^{-p} l_s^{-(p+1)}$ for both the DBI and CS parts, 
we absorb the factor $g_s^2$ originally contained in $2\kappa_{10}^2=2\kappa_{11}^2/(2\pi R_{11})$ by rescaling
$F_{2,4}= g_s F_{2,4}^\mathrm{10d}$ and $e^{-\Phi}= g_s e^{-\Phi_\mathrm{10d}}$, upon which we drop the 10d labels on the fields and
redefine $2\kappa_{10}^2=(2\pi)^7 l_s^8$.

In the coordinates used in \cite{Sakai:2004cn,Sakai:2005yt}, 
\be\label{UandZ}
U=\frac{r^2}{2L},\quad
K(Z)\equiv 1+Z^2=\frac{r^6}{\R^6}=\frac{U^3}{\UKK^3},
\ee
the 10-dimensional metric reads
\be\label{ds210}
ds^2_{10}=\left(\frac{U}{\RSS}\right)^{3/2} \left[\eta_{\mu\nu}dx^\mu dx^\nu
+f(U)dx_4^2\right]+\left(\frac{\RSS}{U}\right)^{3/2}\left[\frac{dU^2}{f(U)}+U^2 d\Omega_4^2 \right]
\ee
with $f(U)=1-(\UKK/U)^3$ and $\RSS\equiv L/2$; the dilaton and 4-form field strength are given by
\be\label{Phibackground}
e^\Phi=g_s(U/\RSS)^{3/4}
\ee
and
\be
F_4=dC_3=\frac{3\RSS^3}{g_s}\omega_4.
\ee
(Here and in the following we stick to the usual normalization of the Ramond-Ramond fields in
the string-theory literature \cite{Becker:2007zj}. In 
\cite{Sakai:2004cn,Sakai:2005yt} these fields are rescaled according to
$C_p=(2\pi)^{p-1}l_s^p C_{p}^{\rm SS}$.)
The 4-form field strength is related to the number $N_c$ of D4 branes by
\be
(2\pi)^2 l_s^3 \int_{S^4} F_4=2\pi N_c,
\ee
which implies $\RSS^3\equiv (L/2)^3=\pi g_s N_c l_s^3$.

The parameters\footnote{Following the notation of \cite{Kruczenski:2003uq,Sakai:2004cn,Sakai:2005yt},
$\gYM^2$ differs by a factor two from the usual particle physics convention so that $\alpha_s=\gYM^2/(2\pi)$.} 
of the dual boundary theory which upon dimensional reduction through the circle $S_\tau$ with radius $\MKK^{-1}$
becomes pure 3+1-dimensional Yang-Mills theory
\be
\mathcal L=-\frac{1}{2\gYM^2}\Tr |\FYM|^2+\frac{\theta}{8\pi^2}\Tr \FYM\wedge\FYM,
\ee
can be identified from the UV limit $U\to\infty$ of the D4 brane action
\bea
S^{\rm D4}&=&-\mu_4 \Tr\int d^4x d\tau e^{-\Phi}\sqrt{-g_{(5)}}(\mathbf1 + \frac12 (2\pi\alpha')^2 |\FYM|^2+\ldots)\nonumber\\
&&+
\mu_4(2\pi\alpha')^2 \int C_1 \wedge\FYM\wedge\FYM + \ldots,
\eea
with $\mu_4=(2\pi)^{-4}l_s^{-5}$ and $\alpha'=l_s^2$.
This gives
\be
\gYM^2=2\pi g_s l_s \MKK,\quad \theta + 2\pi k = \frac{1}{l_s}\int_{S_\tau} C_1
\ee
with $k$ an integer.

The Witten background reviewed above has a vanishing one-form field $C_{\tau}$
and thus corresponds to the theory with vanishing $\theta$ parameter,
or sufficiently small $\theta/N_c$ such that the backreaction on the background
can be neglected, in the $k=0$ branch. (The fully backreacted case has been worked out in Ref.~\cite{Bigazzi:2015bna}.)

The holographic dictionary thus relates non-normalizable
fluctuations of $C_{\tau}$ to a local, possibly $x$-dependent
$\theta$ parameter. Normalizable modes are instead interpreted as
pseudoscalar ($J^{PC}=0^{-+}$) glueball excitations \cite{Brower:2000rp}.

The relevant quadratic action for these
fields is
\begin{align}
S_\mathrm{R}\supset & -\frac{1}{4\kappa_{10}^{2}}\int\d^{10}x\sqrt{-g}\left(\left|F_{2}\right|^{2}\right)\nonumber \\
= & -\frac{2\pi R_4 V_4}{64\kappa_{10}^{2}}\int\d^{4}x
\int_{\rKK}^\infty dr\left(\frac{r^{3}L}{f(r)}\eta^{\mu\nu}\partial_{\mu}C_{\tau}\partial_{\nu}C_{\tau}+\frac{r^{7}}{L^{3}}\left(\partial_{r}C_{\tau}\right)^{2}\right).
\label{eq:10dRamondCt}
\end{align}
The resulting field equations have
the non-normalizable solution 
\begin{align}\label{Ctheta}
C_{\tau}^{\left(0\right)}= & \frac{l_s}{2\pi R_4} f(r)\vartheta(x) \qquad\text{for}\; \Box\,\vartheta(x)=0.
\end{align}
In the Witten model of pure Yang-Mills theory, $\vartheta=\theta$, 
but since this connection will be modified when quarks are included,
we have providently introduced a different symbol.

Inserting (\ref{Ctheta}) in the action gives
\be\label{chig}
S_\mathrm{R}=-\frac{\chi_g}{2} \int d^4x\, \theta^2, \quad \chi_g=\frac{\lambda^3 \MKK^4}{4(3\pi)^6},
\ee
with $\chi_g$ being the topological susceptibility.

The normalizable solutions will be expanded in mass eigenfunctions,
with radial eigenvalue equations
\begin{align}
\partial_{r}\left(\frac{r^{7}}{L^{3}}\partial_{r}C_{\tau}^{(2)}\right)+\frac{r^{3}L}{f(r)}M_G^{2}C_{\tau}^{(2)} & =0,\label{eq:CtEom}
\end{align}
subject to 
boundary conditions $C_{\tau}^{(2)}(r=\infty)=C_{\tau}^{(2)}(r=\rKK)=0$, $\partial_r C_{\tau}^{(2)}(r=\rKK)\not=0$. 
This determines the mass of the lightest pseudoscalar glueball as $M_G=1.885\ldots\times \MKK$.

\subsection{Inclusion of quarks and $U(1)_A$ anomaly}

Sakai and Sugimoto have extended the Witten model by introducing left and right handed chiral quarks
through $N_f$ pairs of D8 and $\overline{\text{D8}}$ probe branes localized at separate points on the circle $S_\tau$ at the holographic boundary.
Chiral symmetry breaking $U(N_f)_L\times U(N_f)_R\to U(N_f)_{L+R}$ emerges from the fact that
the D8 and $\overline{\text{D8}}$ branes have to join in the cigar-shaped background geometry.
Choosing antipodal points on $S_\tau$ leads to embedding functions with constant $x_4\equiv \tau$
and a joining of the branes at the tip of the cigar at $\rKK$ (or $\UKK$). In this case one can extend the coordinate
$Z$ introduced in (\ref{UandZ}) to the range
$-\infty\ldots\infty$ in order to cover the radial extent of joined D8 and $\overline{\text{D8}}$ branes.

The action for these flavor branes is given by $S^\mathrm{D8}=S_\mathrm{DBI}^\mathrm{D8}+S_\mathrm{CS}^\mathrm{D8}$ with
\be
S_\mathrm{DBI}^\mathrm{D8}=-\mu_8 \int_\mathrm{D8} e^{-\Phi}\, \widetilde{\text{Tr}}
\sqrt{-\det\left(g_{(9)}+2\pi\alpha'\mathcal F+B_2\right)},
\ee
where $\widetilde{\text{Tr}}$ denotes symmetrized trace, $\mu_8=(2\pi)^{-8} l_s^{-(9)}$, and
$\mathcal F=d\mathcal A+\mathcal A\wedge \mathcal A$ is the field strength tensor for the nonabelian flavor gauge fields living on the D8-branes.
($B_2$ is the bulk Kalb-Ramond 2-form field; 
its normalizable modes contain the pseudovector ($1^{+-}$) glueballs of the dual gauge theory \cite{Brower:2000rp,Brunner:2018wbv}.)

The even and odd radial mode functions of $\mathcal A_\mu$ are associated with
the towers of vector and axial-vector meson fields, with the lowest mass eigenvalue ($m_{v_1}^2\approx 0.669\MKK^2$) being
identified as the $\rho$ meson mass such that $\MKK=949\,\text{MeV}$.

The massless pseudoscalar Goldstone bosons are described by
\be
U(x)=\mathrm P\,\exp i\left(\int_{-\infty}^\infty dZ A_Z(Z,x)\right)=\exp\left({i\Pi^a\lambda^a/f_\pi}\right),
\ee
where $f_\pi^2=\frac1{54\pi^4}\lambda N_c\MKK^2$ and $\lambda^a$ are Gell-Mann matrices supplemented
by $\lambda^0=(N_f/2)^{-1/2}\mathbf{1}$. Setting $f_\pi=92.4\text{MeV}$ fixes
$\lambda\approx 16.63$ for $N_c=3$.
A smaller value of about 12.55 would be found by matching instead the large-$N$ lattice
result \cite{Bali:2013kia} for the string tension. As in \cite{Brunner:2015oqa} we shall
consider the range $\lambda=16.63\ldots12.55$ in order to obtain a theoretical error band for our quantitative
predictions. 

The 9-dimensional Chern-Simons term of the flavor brane is given by
\begin{align}\label{SCS}
S_\mathrm{CS}^\mathrm{D8}= & \mu_{8}\sum_{q}\int_\mathrm{D8}\sqrt{\hat A(\mathcal R)}\;\text{Tr}\,\exp\left(2\pi\alpha'\mathcal{F}+B\right)\wedge C_{q},
\end{align}
where the so-called A-roof genus factor involves $\hat A(\mathcal R)=1+\frac{1}{192\pi^2}\Tr \mathcal R\wedge\mathcal R+\ldots$
with $\mathcal R^{MN}=\frac12 R_{KL}{}^{MN}dx^K dx^L$.
The term involving $C_3$ contains the Wess-Zumino-Witten term of the dual gauge theory because $F_4=dC_3$ is nonzero in the background \cite{Sakai:2004cn,Lau:2016dxk}.

The Chern-Simons term is also involved in the $U(1)_A$ anomaly of the dual gauge theory, because the $C_7$ term modifies
the equations of motion of the $C_1$ field which is responsible for the $\theta$ parameter.
Using Hodge duality, $dC_7=F_8=\star F_2$, and integrating by parts, one finds that
\begin{align}
S_\mathrm{CS}&\supset  \mu_{8}2\pi\alpha^{\prime}\int_\mathrm{D8}\text{Tr}\left(\mathcal{F}_{2}\wedge C_{7}\right)\nonumber \\
&=  \mu_{8}2\pi\alpha^{\prime}\int\text{Tr}\left(\mathcal{A}_{1}\right)\wedge\star F_{2}\wedge\omega_{\tau}\nonumber \\
&=  \mu_{8}2\pi\alpha^{\prime}\int\d x^{10}\sqrt{\left|g_{10}\right|}\left[\delta\left(\tau\right)+\delta\left(\tau-\pi\right)\right]\nonumber \\
 &\qquad \times\left(\text{Tr}\left(\mathcal{A}_{r}\right)g^{rr}g^{\tau\tau}\partial_{r}C_{\tau}+g^{\mu\nu}g^{\tau\tau}\text{Tr}\left(\mathcal{A}_{\mu}\right)\partial_{\nu}C_{\tau}\right),\label{eq:10dCS}
\end{align}
where $\omega_{\tau}=\left[\delta\left(\tau\right)+\delta\left(\tau-\pi\right)\right]d\tau$ has been introduced
to extend the integration to the entire bulk spacetime.
Thus the flavor probe branes induce a mixing term of the Abelian part of the flavor gauge field $\mathcal{\hat{A}}:=N_f^{-1}\Tr\,\mathcal A$ living
on the D8 brane with the field $C_{\tau}$. The
linear equations of motion for $C_{\tau}$ get an additional term
localized on the D8 brane, 
\bea
\frac{1}{32\kappa_{10}^{2}}\partial_{r}\left(\frac{r^{7}}{L^{3}}\partial_{r}C_{\tau}\right)&=&
2\pi\alpha^{\prime}\mu_{8}\partial_{r}\left(\sqrt{-g}g^{rr}g^{\tau\tau}\,\Tr(\mathcal{A}_r)\right)\left(\delta\left(\tau\right)+\delta\left(\tau-\pi\right)\right)
\nonumber\\
&=&2\pi\alpha^{\prime}\mu_{8}\partial_{r}\left(\frac{r^{7}}{2^{4}L^{3}} N_{f}\mathcal{\hat{A}}_{r}\right)\left(\delta\left(\tau\right)+\delta\left(\tau-\pi\right)\right),
\eea
which we solve by introducing the localized
fluctuation $C_{\tau}^{\left(\delta\right)}$ according to 
\be
\partial_{r}C_{\tau}^{\left(\delta\right)}= 4\pi\alpha^{\prime}\kappa_{10}^{2}\mu_{8}N_{f}\mathcal{\hat{A}}_{r}\left(\delta\left(\tau\right)+\delta\left(\tau-\pi\right)\right),\label{eq:ctdeltaeom}
\ee
where we have assumed $\Box C_\tau^{\left(\delta\right)}=0$ with respect to Minkowski space coordinates
and $\partial_\mu \hat{\mathcal{A}}^\mu=0$.
Switching to $Z^2=(r/\R)^6-1$ and with 
\begin{align}
\mathcal{\hat{A}}_{Z}(Z,x) 
=  \frac{\sqrt{2}}{\sqrt{N_{f}}\pi f_{\pi}}\frac{1}{1+Z^2}\eta_{0}(x),
\end{align}
we thus obtain
\begin{align}\label{Cdelta}
C_{\tau}^{\left(\delta\right)}= & \int_{0}^{Z}dZ\,\partial_{Z}C_{\tau}^{\left(\delta\right)}\nonumber \\
= & 4\pi\alpha^{\prime}\kappa_{10}^{2}\mu_{8}\sqrt{N_{f}}\frac{\sqrt{2}}{\pi f_{\pi}}\arctan(Z)\eta_{0}(x)\left[\delta\left(\tau\right)+\delta\left(\tau-\pi\right)\right].
\end{align}

The ``anomalous'' part $C_{\tau}^{\left(\delta\right)}$ contributes to the $\theta$ parameter,
\bea
\theta &= & l_s^{-1}\int_{S_\tau} C_1 = l_s^{-1}\int_{\rm cigar} F_2 = 
l_s^{-1}\int\d r\d\tau\,\partial_{r}(C_{\tau}^{\left(0\right)} + C_{\tau}^{\left(\delta\right)})\nonumber\\
&= & \vartheta(x)+\frac{2\pi\alpha^{\prime}4\kappa_{10}^{2}\mu_{8}\sqrt{N_{f}}}{\sqrt{2}f_{\pi} l_s}\eta_{0}(x)
=\vartheta(x)+\frac{\sqrt{2N_{f}}}{f_{\pi}}\eta_{0}(x).\label{eq:thteReinterpretation}
\eea
In the presence of flavor branes, the $\theta$ parameter is therefore 
no longer given by
$\vartheta$ alone, but also involves $\eta_0$.

In the remainder of this work we will set the $\theta$ parameter to 0,
which corresponds to non-vanishing $\vartheta(x)$ according
to
\begin{equation}
\vartheta(x)=-\frac{\sqrt{2N_{f}}}{f_{\pi}}\eta_{0}(x) .
\end{equation}
The meson field $\eta_{0}$ therefore also appears in the non-normalizable
mode $C_{\tau}^{\left(0\right)}$,
which is a fundamental ingredient in the realization of the Witten-Veneziano
mechanism in the WSS model.

Note that, in the chiral case, a constant $\theta$ can be absorbed simply in a field redefinition
$\eta_0\to \eta_0+f_\pi \theta/\sqrt{2N_f}$, since only derivatives of $\eta_0$ appear in the effective
action produced by $S_\mathrm{DBI}^\mathrm{D8}$. However, introducing mass terms for quarks either through world-sheet instantons
or nonnormalizable modes of bifundamental fields corresponding to open-string tachyons \cite{0708.2839,Dhar:2008um,Aharony:2008an,Hashimoto:2008sr,McNees:2008km,Niarchos:2010ki}
produces the additional term
\be\label{LMU}
\mathcal L_m^{\mathcal M} \propto\int d^4x \,\Tr\left(\mathcal M\,U(x)+h.c.\right),
\ee
where such a redefinition changes the phase of the quark mass matrix $\mathcal M={\rm diag}(m_u,m_d,m_s)$
according to $\mathcal M\to \mathcal M e^{i\theta/N_f}$.

Both, in the chiral limit and in the case with nonzero quark masses, the singlet $\eta_0$ receives
an extra mass term that is determined by the topological susceptibility 
of pure Yang-Mills theory obtained in (\ref{chig}),
\be
m_0^2=\frac{2N_f}{f_\pi^2}\chi_g,
\ee
in accordance with the Witten-Veneziano formula \cite{Witten:1979vv,Veneziano:1979ec}.

\section{Pseudoscalar glueball-meson mixing}\label{sec:mixing}

With the additional term $C_{\tau}^{\left(\delta\right)}\propto\eta_0$ we have
solved the anomalous equations of motion.\footnote{As stated after (\ref{eq:ctdeltaeom}),
we had to assume $\Box C_{\tau}^{\left(\delta\right)}=0$ with respect to Minkowski coordinates.
With $\eta_0$ picking up the mass $m_0^2$, this assumption is violated, but only
at higher order in $N_f$: 
$\Box C_{\tau}^{\left(\delta\right)}\sim N_f^{1/2}\Box\eta_0\propto N_f^{3/2}$.
(With nonzero quark masses, this seems safe as long as their contributions to the masses
of the pseudoscalar mesons are much smaller than $m_0$.)}
Now we will determine the
field redefinitions that are necessary to obtain a diagonal
action for the Minkowski space fields $\eta_{0}$ and $\tilde G$. In doing this
we encounter divergent terms proportional to $\delta(0)$ in analogy to
the Ho\v{r}ava-Witten calculation \cite{Horava:1996ma}. By adding to the Lagrangian terms beyond
the probe approximation one would presumably be able to cancel these
divergences. In the following we will however just drop terms proportional
to $\delta(0)$, i.e., the $(C_{\tau}^{\left(\delta\right)})^{2}$
term in $S_\mathrm{R}$ and the $C_{\tau}^{\left(\delta\right)}$ term
in $S_\mathrm{CS}$.

Let us start with the effective kinetic terms coming from 
the first part of (\ref{eq:10dRamondCt}),
\be
S_\mathrm{R}^{\left(\text{kin}\right)}=-\frac{\pi^3}{12\kappa_{10}^2\MKK}\int_{\rKK}^\infty dr \int d^4x 
\frac{r^{3}L}{f(r)}\eta^{\mu\nu}\partial_{\mu}C_{\tau}\partial_{\nu}C_{\tau}.
\ee
With $C_\tau=C_{\tau}^{\left(0\right)}+C_{\tau}^{\left(\delta\right)}+C_\tau^{(2)}$
and $C_\tau^{(2)}(r,x)\propto \tilde G(x)$
we write them as 
\begin{align}\label{SRkin}
S_\mathrm{R}^{\left(\text{kin}\right)}= & \int\d^{4}x\left(\1\partial_{\mu}\eta_{0}\partial^{\mu}\eta_{0}+\2\partial_{\mu}\eta_{0}\partial^{\mu}\tilde G
-\frac12\partial_{\mu}\tilde G\partial^{\mu}\tilde G\right),
\end{align}
where we have fixed the normalization of the radial mode functions in $C_\tau^{(2)}$ by requiring
\be
\frac{\pi^3}{12\kappa_{10}^2\MKK}\int_{\rKK}^\infty dr \frac{r^{3}L}{f(r)}
\left(C_{\tau}^{(2)}/\tilde G\right)^{2} = \frac12.
\ee
For the constant $\1$ which corresponds to a wave function renormalization of $\eta_0$ we obtain
\begin{align}
\1= & 
-\frac{\pi^3}{12\kappa_{10}^2\MKK}\int_{\rKK}^\infty dr \frac{r^{3}L}{f(r)}
\left(C_{\tau}^{\left(0\right)}+2C_{\tau}^{\left(\delta\right)}\right)C_{\tau}^{\left(0\right)}/\eta_{0}^{2}
=: \frac{N_{f}}{N_{c}}\9,
\end{align}
where $\9$ is divergent, since it is obtained by integrating non-normalizable modes,
but $\1$ is suppressed by a factor $N_c/N_f$ compared to the kinetic term for $\eta_0$ contained in $S_\mathrm{DBI}^\mathrm{D8}$.
The constant $\2$ which is associated with a kinetic mixing of $\eta_0$ and pseudoscalar glueball modes
is finite and given by
\bea
\2&=&  
-\frac{\pi^3}{12\kappa_{10}^2\MKK}\int_{\rKK}^\infty dr \frac{r^{3}L}{f(r)}
\left(C_{\tau}^{\left(0\right)}+C_{\tau}^{\left(\delta\right)}\right)C_{\tau}^{(2)}/\left(\eta_{0}\tilde G\right) \nonumber \\
&=&  0.011180\ldots 
\sqrt{\frac{N_{f}}{N_{c}}}\lambda\label{eq:c2}
\eea
for the lowest pseudoscalar glueball. (The next-to-lightest, excited pseudoscalar glueball has
$\2=-0.014314\ldots \sqrt{N_f/N_c}\lambda$.)

The remaining terms from the background action (\ref{eq:10dRamondCt})
and the CS action (\ref{eq:10dCS})
\begin{align}
S_\mathrm{R}^{\left(\text{mass}\right)}= & -\frac{1}{4\kappa_{10}^{2}}\int\d^{10}x\sqrt{g_{S^{4}}}\frac{1}{2^{4}}\frac{r^{7}}{L^{3}}\left(\partial_{r}C_{\tau}\right)^{2}\nonumber \\
 & +\mu_{8}2\pi\alpha^{\prime}\int\d^{10} x\left(\delta\left(\tau\right)+\delta\left(\tau-\pi\right)\right)\cdot\sqrt{\left|g_{10}\right|}\text{Tr}\left(\mathcal{A}_{r}\right)g^{rr}g^{\tau\tau}\partial_{r}C_{\tau}
\end{align}
give the effective mass terms
\begin{align}
S_\mathrm{R}^{\left(\text{mass}\right)}= & -\frac12 m_0^2 \eta_{0}^{2}-\frac12 M_G^2 \tilde G^{2}
+\zeta_3 \eta_0 \tilde G,
\end{align}
with the values already determined above,
\be
m_0^2=  
\frac{\pi V_4}{\kappa_{10}^2 \MKK}\int_{\rKK}^\infty dr
\frac{1}{2^{4}}\frac{r^{7}}{L^{3}}\left(\partial_{r}C_{\tau}^{\left(0\right)}/\eta_0\right)^{2}
=  \frac{\lambda^{2}N_{f}}{27\pi^{2}N_{c}}M_{\kk}^{2} 
\ee
and (for the lightest pseudoscalar glueball mode)
\be
M_G^2=  
\frac{\pi V_4}{\kappa_{10}^2 \MKK}\int_{\rKK}^\infty dr
\frac{1}{2^{4}}\frac{r^{7}}{L^{3}}\left(\partial_{r}C_{\tau}^{(2)}/\tilde G\right)^{2}
=  3.5539 M_{\kk}^{2}.\label{eq:d2}
\ee
The terms involving $\partial_r C_{\tau}^{\left(\delta\right)}\partial_r[C_{\tau}^{\left(0\right)}+C_{\tau}^{(2)}]$
cancel by the equation of motion (\ref{eq:ctdeltaeom}) with the CS term. 
The mass mixing term $\eta_0 \tilde G$ vanishes because
\be\label{zeta3}
\zeta_3 \propto \int_{\rKK}^\infty dr\,
r^{7} \partial_{r}C_{\tau}^{(0)}\partial_{r}C_{\tau}^{(2)} 
 \propto  
 \int_{\rKK}^\infty dr\,
\partial_{r}C_{\tau}^{(2)} =0,
\ee
as already pointed out in \cite{Brunner:2016ygk}.

However, the term involving $C_{\tau}^{\left(\delta\right)}C_{\tau}^{(2)}$ in (\ref{eq:c2}) has
produced a nontrivial \emph{kinetic} mixing term of order $(N_f/N_c)^{1/2}$. 
To get rid of such mixing terms one has to perform a non-unitary field redefinition. In the present case,
one has to make the substitutions
\begin{align}\label{c12fieldrefs}
\eta_{0}\rightarrow & \left(1+\1\right){\eta}_{0}+\2 {\tilde G},\nonumber \\
\tilde G\rightarrow & \left(1+\frac12 \2^2\right){\tilde G},
\end{align}
which yield
\bea\label{Lbilinunmixed}
\mathcal L^{\text{bilin.}}_{\eta_0,\tilde G}&=&
-\frac12 (\partial_\mu \eta_0)^2-\frac12 (\partial_\mu \tilde G)^2
+\partial_\mu \eta_0 \partial^\mu \tilde G \left(2 \1 \2+\frac12 \2^3 \right)\nonumber\\
&&-\frac12 M_G^2 (1+\2^2) \tilde G^2-\frac12 m_0^2 \eta_0^2 - m_0^2 \2 \eta_0 \tilde G + O(N_f^2/N_c^2) \\
&=&
-\frac12 (\partial_\mu \eta_0)^2-\frac12 (\partial_\mu \tilde G)^2
-\frac12 M_G^2 (1+\2^2) \tilde G^2-\frac12 m_0^2 \eta_0^2  + O(N_f^{3/2}/N_c^{3/2}),\nonumber
\eea
where we have taken into account that $\2\propto \sqrt{N_f/N_c}$ while $\1,m_0^2 \propto N_f/N_c$.
The field redefinitions (\ref{c12fieldrefs}) thus diagonalize the bilinear terms up to and including
order $N_f/N_c$, yielding also a small positive contribution to the pseudoscalar glueball mass term,
\be\label{MGcorr}
M_G^2=(1789.0 \,\text{MeV})^2\to \left(1+(0.01118 \,\lambda)^2 N_f/N_c\right)M_G^2=(1819.7\ldots1806.5 \,\text{MeV})^2
\ee
for $\lambda=16.63\ldots12.55$.

Dropping all terms of order $(N_f/N_c)^{3/2}$ and higher, we see that the divergent coefficient $\1$
can be ignored and that 
(\ref{c12fieldrefs}) reduces to
\begin{align}
\eta_{0}\rightarrow & \,\eta_0+\2\,\tilde G= {\eta}_{0}+0.01118\sqrt{\frac{N_{f}}{N_{c}}}\lambda\,{\tilde G},\nonumber \\
\tilde G\rightarrow & \,{\tilde G}.
\label{eq:leadingMixing}
\end{align}

Note that this is very different from the mixing scenario proposed by Rosenzweig et al.\ in \cite{Rosenzweig:1981cu,Rosenzweig:1982cb}
and also considered in \cite{Mathieu:2009sg}. In that scenario it is assumed that the chiral anomaly is not saturated by $\eta_0$ alone, but
only together with the pseudoscalar glueball field\footnote{According to \cite{Rosenzweig:1981cu}, this additional
field could however also be interpreted as a radial excitation of $\eta'$.}. 
This leads to a nondiagonal mass matrix for $\eta_0$ and the pseudoscalar glueball
which can be diagonalized by an orthogonal matrix. The field definition (\ref{c12fieldrefs}), on the other hand, is nonunitary.
It is needed to remove the kinetic mixing term (\ref{SRkin}), but originally there is no mass mixing term because of (\ref{zeta3});
a mass mixing term appears after the field redefinition in (\ref{Lbilinunmixed}), but only at the order $N_f^{3/2}/N_c^{3/2}$ which is
beyond the probe approximation.\footnote{In \cite{Escribano:2008rq,Mathieu:2009sg,Ambrosino:2009sc} a unitary mixing of $\eta'$ and pseudoscalar glueballs
has been considered from a phenomenological point of view, which has an important impact on the determination of the pseudoscalar mixing angle $\theta_P$
when there is significant $\eta'$-$\tilde G$ mixing. By contrast, the nonunitary transformation (\ref{eq:leadingMixing}) that we obtained
does not have this effect (to leading order).}

In Ref.~\cite{Brunner:2016ygk}, where (in view of the vanishing mass mixing term) the interactions of the unmixed pseudoscalar glueball
were considered, it was found that those are given either by pairs of pseudoscalar glueballs
interacting with scalar or tensor glueballs or by a vertex connecting
a pseudoscalar glueball with $\eta_0$ and a scalar glueball.
The latter, which is relevant for the decay of a pseudoscalar glueball, is due to the fact
that the integral (\ref{zeta3}) for the $\eta_0\tilde G$ mass mixing term no longer vanishes
when metric fluctuations dual to a scalar glueball are inserted.
Assuming that this is the dominant vertex, a very narrow pseudoscalar glueball was
predicted whose decays had to involve $\eta$ or $\eta'$ together with the decay products
of a scalar glueball.

Through the shift (\ref{eq:leadingMixing}) a pseudoscalar glueball
acquires also all types of vertices that the singlet $\eta_0$ possesses.
In the DBI part of the D8 brane action, such vertices are contained only in terms
involving $\mathcal F$ to fourth and higher power, which are suppressed by
higher powers of $\alpha'$. In the chiral limit, the dominant interactions
come from the CS part of the D8 brane. With quark masses introduced according to
(\ref{LMU}), further interactions arise from (assuming $\mathcal M=\mathcal M^\dagger$)
\be\label{LMUshift}
\Tr(\mathcal M (U+U^\dagger))\to
\Tr(\mathcal M (U+U^\dagger))+i\2 f_\pi^{-1}\sqrt{2/{N_f}}\,\tilde G \, \Tr(\mathcal M (U-U^\dagger)).
\ee

The additional term in (\ref{LMUshift}) also contains a bilinear term involving $\tilde G$ and $\eta_{0,8}$,
\be\label{DeltaLm2}
\Delta \mathcal{L}_m^{(2)}=-\frac{2\2}{\sqrt3} \tilde G \left[
\frac1{\sqrt3} (m_K^2+\frac12 m_\pi^2)\,\eta_0-\sqrt{\frac23}(m_K^2-m_\pi^2)\,\eta_8 \right].
\ee
It implies a further mixing of $\tilde G$ with $\eta_{0,8}$ which turns out to be
negligible as concerns the masses of $\eta$, $\eta'$, and $\tilde G$ when the mass
matrix of the $\tilde G$-$\eta_{0}$-$\eta_8$ sector is diagonalized. Due to the large mass
of the pseudoscalar glueball, the corrections to the masses of $\eta$, $\eta'$, and $\tilde G$
are only about $-0.015\%$, $-0.008\%$, and $+0.09\%$, respectively, and thus can be safely neglected.
Also the vertices induced by the kinetic mixing of $\tilde G$ and $\eta_0$ receive only small
corrections of the order $\2 m_K^2/m_G^2\ll \2$. However, the small mixing of $\tilde G$ with $\eta_8$
gives rise to new types of vertices from the DBI action.

In the next section we shall consider the consequences of all these additional interactions
in turn.

\section{Pseudoscalar glueball decay modes}\label{sec:decay}

\subsection{Decay into two vector mesons}

The dominant decay channel of the pseudoscalar glueball
mode turns out to be a decay into two vector mesons. This decay arises from the Chern-Simons term
and is mediated
by the mixing of $\eta_{0}$ with the glueball $\tilde G$ according to (\ref{eq:leadingMixing}).
Before the field redefinition we have the interaction term
\begin{align}
S_\mathrm{CS}^\mathrm{D8} &\supset  \mu_{8}\frac{\left(2\pi\alpha'\right)^{3}}{3!}\int\text{Tr}\left(\mathcal{F}\wedge \mathcal{F}\wedge \mathcal{F}\wedge C_{3}\right)\nonumber \\
&\supset  \mu_{8}\frac{\left(2\pi\alpha'\right)^{3}}{2g_{s}}L^{3}\pi^{2}\epsilon^{\mu\nu\rho\sigma}\int\text{Tr}\left(\hat{\mathcal{A}}_{Z}\partial_{\mu}\mathcal{A}_{\nu}\partial_{\rho}\mathcal{A}_{\sigma}\right),
\end{align}
with $\hat{\mathcal{A}}_{Z}\propto\eta_0$,
which after the field redefinition yields the term
\begin{align}\label{SGtvv}
S_\mathrm{CS}^\mathrm{D8} &\supset  k_{1}\tilde G\epsilon^{\mu\nu\rho\sigma}\text{Tr}\partial_{\mu}v_{\nu}\partial_{\rho}v_{\sigma},
\end{align}
with the coupling constant
\begin{align}
k_{1} &=  
\2\,\frac{1}{\sqrt{N_{f}}}U_{\kk}\mu_{8}\frac{\left(2\pi\alpha'\right)^{3}}{2g_{s}}L^{3}\pi^{2}\int\d Z\phi_{0}\psi_{1}^{2}\nonumber \\
&= \2\times  877.39 \,\lambda^{-3/2} N_f^{-1/2} N_c^{-1/2} \MKK^{-1} \nonumber\\
&=  9.8092
\,M_{\kk}^{-1}N_{c}^{-1}\lambda^{-\frac{1}{2}},
\end{align}
for the lowest vector meson mode $\psi_1$. For the lowest axial vector mesons with radial mode function $\psi_2$ 
with mass square eigenvalue $m_{a_1}^2\approx 1.57\MKK^2 \approx (1190 \text{MeV})^2$, which is quite close to the experimental value of 1230 MeV of the lightest
$a_1$ axial vector meson,\footnote{In table \ref{tab:2v} we ignore the slightly
heavier $f_1$ and $K_1$ axial vector mesons, which would start contributing significantly only for $M_G$ well
above 2600 MeV.}
we obtain instead $k_2=4.8888 \,M_{\kk}^{-1}N_{c}^{-1}\lambda^{-\frac{1}{2}}$.

Performing the polarization sums we obtain the amplitude squared
\begin{align}
\sum_{\epsilon_1,\epsilon_2}\left|\mathcal{M}(\tilde G\to vv)\right|^{2} &=  8k_{1}^{2}M^{2}\left(\frac{M^{2}}{4}-m^{2}\right),
\end{align}
where $M$ and $m$ are the masses of $\tilde G$ and $v$ (and similarly for the heavier vector and axial vector modes when $M$
is large enough).

The partial widths for decays into the various vector mesons are listed in table \ref{tab:2v}
for $\lambda=16.63\ldots12.55$ and glueball mass $M_G$ given alternatively by the WSS result (\ref{MGcorr})
and by an (admittedly speculative)
extrapolation to the lattice prediction of 2600 MeV, where 
we have assumed that the mixing parameter $\2$ and thus $k_1$ (which has inverse mass dimension)
scales like $M_G^{-1}$ when the glueball
mass is raised. (Keeping the mixing parameter as is, the widths for a 2600 MeV glueball would all be
a factor of about 2 larger.)
Also with the assumed reduction of $k_1$, a 2600 MeV pseudoscalar glueball
is thus projected to be a rather broad resonance. 

\begin{table}[tbp]
\centering
\begin{tabular}{|l|c|c|}
\hline
 & $\Gamma_p$($M_G$=1813$\pm$7MeV) & $\Gamma_p$($M_G$=2600MeV) \\
\hline
$\rho\rho$ & 36.8\ldots45.0 & 190\ldots248 \\
$\omega\omega$ & 11.4\ldots13.8 & 62\ldots81 \\
$K^*\bar{K}^*$ & 2.7\ldots1.8 & 189\ldots246 \\
$\phi\phi$ & - & 29\ldots38 \\
$a_1 a_1$ & - & 3.1\ldots4.0 \\
\hline
$\sum_{vv}$ & 51\ldots61 & 473\ldots618 \\
\hline
\end{tabular}
\caption{\label{tab:2v} Partial decay widths in MeV for the decay of a pseudoscalar glueball in two vector mesons
with $\lambda=16.63\ldots12.55$ and $M_G$ given by the WSS result (\ref{MGcorr}) and also when
extrapolated to the lattice prediction of 2600 MeV with reduced mixing (see text).}
\end{table}

\subsection{Decay into three pseudoscalar mesons}

In Ref.~\cite{Gounaris:1988rp}, Gounaris and Neufeld have proposed that a pure pseudoscalar glueball decays predominantly through
an interaction Lagrangian involving $i\tilde G\, \Tr(M(U-U^\dagger))$ in order to explain that the 
pseudoscalar glueball candidate at the time, $\iota(1460)$,\footnote{This is now listed as two states
$\eta(1405)$ and $\eta(1475)$ by the Data Particle Group \cite{PDG18}, of which 
the lighter one is still considered occasionally
as a glueball candidate \cite{Masoni:2006rz} in view of its supernumerary nature
with respect to the quark model, while others question
the existence of two separate states and argue for a single $\eta(1440)$ \cite{Bugg:2009cf}.} seemed to decay mainly into $K\bar K\pi$.
Decays into $\eta\pi\pi$ would thus be suppressed by a factor $(m_\pi/m_K)^4$.

In sect.~\ref{sec:mixing} we have found that such an interaction Lagrangian is in fact generated in the WSS model
with nonzero quark masses
by the kinetic mixing of $\eta_0$ and pseudoscalar glueball modes.
Explicitly, it reads
\be\label{DeltaLm}
\Delta\mathcal{L}_m=i\2 \frac{f_\pi}{2\sqrt{2N_f}} \,\tilde{G}\, \Tr \bar{\mathcal M}(U-U^\dagger),
\ee
with
\be
\bar{\mathcal M} = \text{diag} \left( m_\pi^2,\, m_\pi^2,\, 2 m_K^2-m_\pi^2 \right)
\ee
in the isospin symmetric case $m_u=m_d$.
This contains vertices of the pseudoscalar glueball with $K\bar K\pi,K\bar K\eta(')$, and $\eta(')^3$
of the order of $\2 m_K^2/f_\pi^2$, and vertices with $\eta(')\pi\pi$ proportional to $\2 m_\pi^2/f_\pi^2$.

In table \ref{tab:3m} we list the resulting partial decay widths for a pseudoscalar glueball
with mass $M=1813\pm7$ MeV according to (\ref{MGcorr}) and, as above,
with mass extrapolated to 2600 MeV as predicted by quenched lattice QCD.
\begin{table}[tbp]
\centering
\begin{tabular}{|l|c|c|}
\hline
 & $\Gamma_p$($M_G$=1813$\pm$7MeV) & $\Gamma_p$($M_G$=2600MeV) \\
\hline
$K\bar K\pi$ (w\!/\!o $KK^*$) & 0.398\ldots0.387 & 0.558\ldots 0.550\\
\hline
$K\bar K\eta$  (w\!/\!o $f_0(1710)\eta$ etc.) & 0.0263\ldots 0.0064 & 0.1092\ldots 0.0285\\
$K\bar K\eta'$ & - & 0.3303\ldots 0.3570\\
$\pi\pi\eta$ & 0.0048\ldots  0.0061 & 0.0049\ldots 0.0061\\
$\pi\pi\eta'$ & 0.0011\ldots  0.0007 & 0.0021\ldots 0.0013\\
$\eta\eta\eta$ & 0.0039\ldots  0.0007 & 0.0337\ldots 0.0064\\
$\eta\eta\eta'$ & - & 0.0564\ldots 0.0277\\
$\eta\eta'\eta'$ & - & 0.0047\ldots 0.0048\\
\hline
$PP\eta^{(}{}'^{)}$ (w\!/\!o $f_0(1710)\eta^{(}{}'^{)}$)
& 0.036\ldots  0.014 & 0.540\ldots 0.431  \\
\hline
$f_0(1710)\eta^{(}{}'{}^{)}\to PP\eta^{(}{}'{}^{)}$ \cite{Brunner:2016ygk} & 0.0068 \ldots 0.015 & 2.5\ldots 4.3 \\
\hline
\end{tabular}
\caption{\label{tab:3m} Partial decay widths in MeV for the decay of a pseudoscalar glueball in three pseudoscalar mesons
resulting from (\ref{DeltaLm}) with $\lambda=16.63\ldots 12.55$. The results for $\tilde G\to K\bar K\pi$ do not 
yet include the decays $\tilde G\to KK^*\to K\bar K\pi$) discussed in sec.~\ref{sec:decPV}, and those for 
$\tilde G\to PP\eta(')$ with $P$ a pseudoscalar meson neglect $\tilde G\to f_0(1710)\eta^{(}{}'^{)}\to PP\eta^{(}{}'{}^{)}$.
For comparison, the decay modes for an unmixed pseudoscalar glueball via
$\tilde G\to f_0(1710)\eta(')$
as obtained in Ref.~\cite{Brunner:2016ygk} 
are evaluated 
in the last entry of the table for the same set of parameters. These latter results are however
incomplete in the mixed case (see text).
}
\end{table}

In chiral perturbation theory, a well-known problem is that the leading-order Lagrangian severely underestimates
the decay $\eta'\to\eta\pi\pi$ whose leading-order amplitude is proportional to $m_\pi^2$, to wit,
$|\mathcal M(\eta'\to\eta\pi\pi)|=|2\sqrt{2}\cos(2\theta_P)-\sin(2\theta_P)|(m_\pi/f_\pi)^2/6
$. This accounts for only about 3\% of the experimental result.
Higher-order terms
in chiral perturbation theory which are not present in the WSS model
have been shown to give contributions which are much larger \cite{Beisert:2002ad,Escribano:2010wt}. By the same token,
the partial decay rates of a pseudoscalar glueball into three pseudoscalar mesons that result from its mixing
with $\eta_0$ can be expected to be strongly underestimated by (\ref{DeltaLm}), in particular the amplitudes for decay into $\eta\pi\pi$,
which are proportional to $m_\pi^2$. The other decay amplitudes are of the order of $m_K^2$, but
since $m_K$ is small compared to $m_G$, formally higher-order contributions could again be much more important.
However, even if we assume that these
partial decay widths are underestimated by an order of magnitude, 
they are still much smaller than the decays into two vector mesons.

In Ref.~\cite{Brunner:2016ygk}, one of us with F.\ Br\"unner has calculated the decays of an unmixed pseudoscalar glueball
into three pseudoscalar mesons in the WSS model, which necessarily involves $\eta_0$ and a scalar glueball.
If the latter is identified with $f_0(1710)$, which decays predominantly into kaons (as is predicted \cite{1504.05815,1510.07605} by the
WSS model when scalar glueballs have no or negligible decay width into $\eta\eta'$),
the dominant channel is $\tilde G\to f_0(1710)\eta\to K\bar K\eta$, whereas
$KK\pi$ decays are excluded. 
With mixing, additional amplitudes for $\tilde G\to PP\eta(')$ arise from
$G_\mathrm{scalar}\eta_0^2$ and $G_\mathrm{scalar,tensor}(\partial_\mu\eta_0)^2$ terms in the DBI action and also from the additional
terms involving $C_\tau^{(\delta)}$ in $S_R$. They 
would have to be included in a complete calculation of $PP\eta(')$ decays.
To give an idea of the magnitude of the contributions from $\tilde G\to f_0(1710)\eta(')\to PP\eta(')$,
in table \ref{tab:3m}
we have included the partial widths of these decay modes as obtained in the unmixed case in \cite{Brunner:2016ygk}.
While such contributions are moderate for the pseudoscalar glueball mass (\ref{MGcorr}), they become
important for $M_G$ extrapolated to the mass predicted by lattice which is well above the threshold for $f_0(1710)+\eta$ decays.
However, also these contributions are always much smaller than those for the decays $\tilde G\to vv$.

\subsection{Decay into one pseudoscalar meson together with one or two vector mesons}\label{sec:decPV}

The very small mixing of a pseudoscalar glueball with $\eta_8$ that is induced by the quark mass term
(\ref{DeltaLm2}) also gives rise to vertices with one pseudoscalar meson and one or two vector mesons
due to the terms involving $\Tr(\partial_\mu \Pi,[V^\mu,\Pi])$ and $\Tr([V_\mu,\Pi][V^\mu,\Pi])$
in the Yang-Mills part of the effective action. When the glueball mass is above the respective threshold,
this gives rise to decay modes
\be
\tilde G\to 
KK^*,\pi K^* K^*,KK^*\rho,KK^*\omega,KK^*\phi,\eta(')K^*K^*,
\ee
where $KK^*$ is short for $\bar{K}K^*,K\bar{K}^*$ etc.
The corresponding amplitudes are proportional to $\2 (m_K^2-m_\pi^2)/m_G^2\sim 10^{-2}$ times 
a factor $g_{\rho\pi\pi}\propto \lambda^{-1/2}N_c^{-1/2}$ or $g_{\rho\rho\pi\pi}\propto \lambda^{-1}N_c^{-1}$.

\begin{table}[tbp]
\centering
\begin{tabular}{|l|c|c|}
\hline
 & $\Gamma_p$($M_G$=1813$\pm$7MeV) & $\Gamma_p$($M_G$=2600MeV) \\
\hline
$KK^*$ & 0.381\ldots 0.288 & 0.302\ldots0.225\\ 
\hline
$\pi K^* K^*$ & - & 0.0113\ldots0.0112\\
$KK^*\rho$ & - & (2.50\ldots2.47)$\times10^{-3}$\\
$KK^*\omega$ & - & (0.799\ldots0.787)$\times10^{-3}$\\
$KK^*\phi$ & - & (0.253\ldots0.249)$\times10^{-3}$\\
$\eta K^*K^*$ & - & (0.097\ldots0.026)$\times10^{-3}$\\
\hline
\end{tabular}
\caption{\label{tab:mixed} Partial decay widths in MeV for the decay of a pseudoscalar glueball into one pseudoscalar meson and one or two vector mesons.}
\end{table}

In table \ref{tab:mixed} the resulting partial decay widths are listed.
While the partial width for a decay of the pseudoscalar glueball into $K^*$, one further vector meson and one pseudoscalar meson
are rather small, 
the partial width for $\tilde G \to KK^*$ (and subsequently $\to KK\pi$) 
turns out to be comparable to that of the direct decay $\tilde G \to KK\pi$ evaluated above.
Since we assume that the latter may be underestimated significantly, we have not carried out
a full calculation of $\tilde G \to KK\pi$ combining coherently both amplitudes.

In summary, we are finding a rich pattern of decays of the pseudoscalar glueball in three or more mesons,
which involve vertices which are proportional to pseudoscalar meson masses. 
The by far dominant decay modes are however given by two vector mesons.

\section{Discussion and outlook}\label{sec:conclusion}

\begin{figure}\centering
\includegraphics[width=0.7\textwidth]{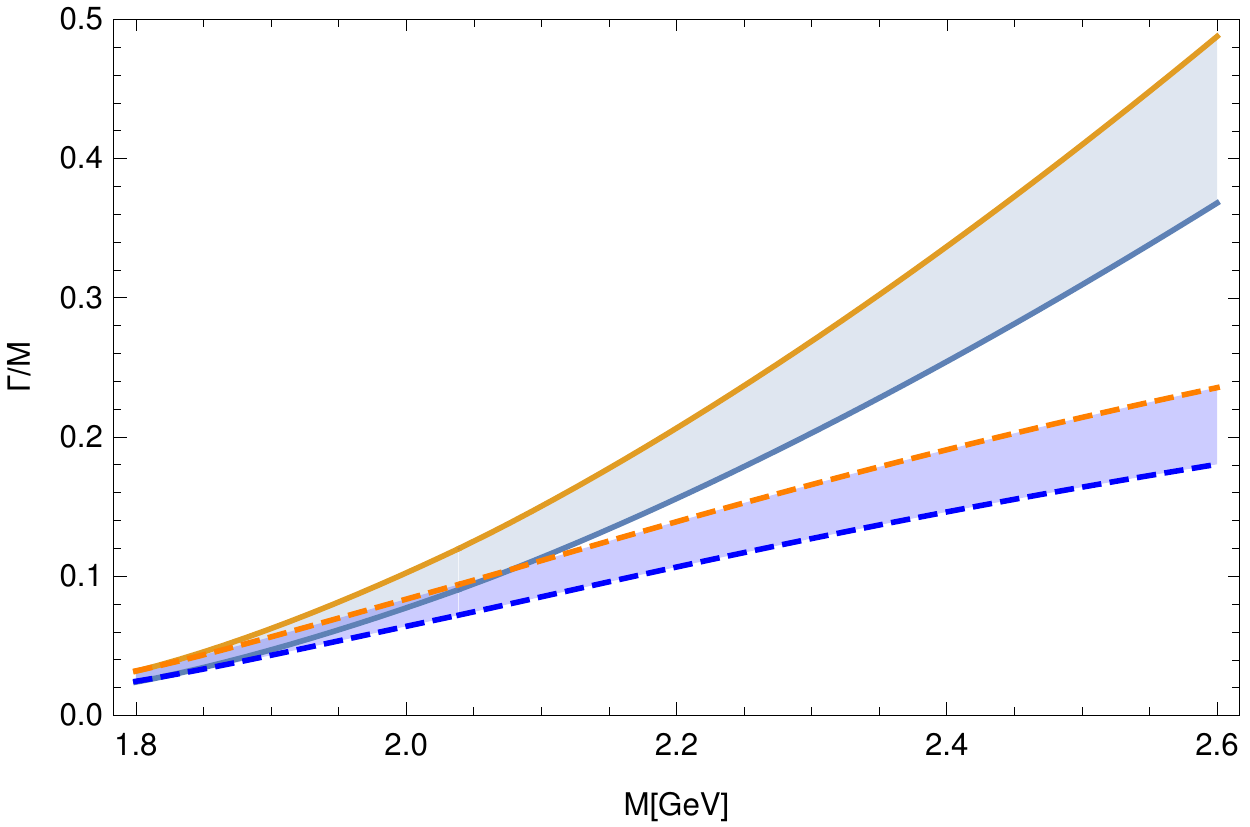}
\caption{Relative pseudoscalar glueball decay width $\Gamma/M$ 
for $\lambda=16.63$ (blue) and 12.55 (orange)
with two different extrapolations to larger masses:
same vertices and mixing (full lines)
or when the mixing parameter $\2$ is assumed to be decreasing like $M_\mathrm{WSS}/M$ (dashed lines).
}
\label{figGPM}
\end{figure}

If the mass of the pseudoscalar glueball is as given by the WSS model, (\ref{MGcorr}), the pseudoscalar glueball is predicted to be
a rather narrow resonance with relative width $\Gamma/M \approx 0.03\ldots 0.04$.
However, the WSS model appears to underestimate significantly the mass of heavier glueballs.
The tensor glueball
comes out as 1487 MeV, while it is predicated as around 2400 MeV by both quenched and unquenched lattice QCD \cite{Morningstar:1999rf,Chen:2005mg,Richards:2010ck,Gregory:2012hu,Sun:2017ipk}.
Extrapolating our results to a pseudoscalar glueball mass
as indicated by (quenched) lattice QCD, 2.6 GeV, the prediction is
instead that of a very broad resonance. In fig.~\ref{figGPM} we display two possible extrapolations,
one with unchanged vertices and mixing, which leads to an extremely large total width,
and the somewhat more moderate scenario (underlying the above tables) where the mixing parameter $\2$ is assumed to be decreasing like $M_\mathrm{WSS}/M$.

The rather large decay width of a pseudoscalar glueball to two vector mesons that we obtained
from the WSS model also implies
a significant coupling to photons because of the vector meson dominance that is inherent in the WSS model \cite{Sakai:2005yt}. 
This also applies to the scalar and tensor glueballs studied
before in \cite{Brunner:2015oqa}. Radiative glueball decays 
will be studied in detail in
forthcoming work, as well as the contribution of glueballs to hadronic light-by-light scattering,
which is of relevance to the theoretical prediction for the muon $g-2$, where holographic QCD
has been argued to provide a promising framework for quantitative predictions \cite{Leutgeb:2019zpq}.

An important implication of the mixing-induced vertices determined in this work is that contrary to the unmixed
case \cite{Brunner:2015oqa} the WSS model predicts sizeable vertices for the production of single pseudoscalar
glueballs in double diffractive scattering. While Regge regime hadronic scattering requires
extensions and extrapolations of the WSS model \cite{Anderson:2014jia,Anderson:2016zon,Hu:2017iix}, the structure of the vertices with Reggeons
and Pomerons are completely determined by the Chern-Simons action and the mixing of the pseudoscalar glueball with $\eta_0$.
For Reggeons, the vertices have the form (\ref{SGtvv}), while Pomeron vertices can be derived from the A-roof genus factor
in (\ref{SCS}) as in \cite{Anderson:2014jia}. This leads to\footnote{%
Up to partial integrations, the tensorial structure of the two terms in (\ref{LPPGt}) is in fact in one-to-one correspondence
to the two couplings $g'_{\T\T\tilde M}$ and $g''_{\T\T\tilde M}$ for two tensor pomerons with one pseudoscalar meson in \cite{Lebiedowicz:2013ika}.} 
\be\label{LPPGt}
\mathcal{L}_{\text{CS}}\supset  \2\tilde G \epsilon^{\mu\nu\rho\sigma}\left[
\kappa_{a}\partial_{\nu}\T_{\alpha\mu}\partial_{\sigma}\T_{\,\rho}^{\alpha}+\kappa_{b}\epsilon^{\mu\nu\rho\sigma}\partial_{\nu}\partial^{\alpha}\T_{\mu\beta}\left(\partial_{\sigma}\partial^{\beta}\T_{\alpha\rho}-\partial_{\sigma}\partial_{\alpha}\T_{\;\rho}^{\beta}\right)\right]
\ee
with $\2\kappa_{a}\MKK$, $\2\kappa_{b}\MKK^3\propto \lambda^{-1/2}N_c^{-2}$.
The results for the unmixed case of ref.\ \cite{Brunner:2015oqa} instead implied that 
central exclusive production of pseudoscalar glueballs required the formation of pairs
$\tilde G \tilde G$, $\eta(')\tilde G$, or $G_\mathrm{scalar,tensor} \tilde G$
with vertices of order $\lambda^{-1}N_c^{-2}$, $N_f^{1/2} N_c^{-5/2}$, or $\lambda^{-1}N_f N_c^{-3}$, respectively.
The possibility of production of single pseudoscalar glueballs of course lowers the threshold significantly.

Similarly, the kinetic mixing of the pseudoscalar glueball with $\eta_0$ allows the production of
single pseudoscalar glueballs in radiative $J/\psi$ decays with rates
of the parametric order $N_c^{-1}$ times that of the decay rate for $J/\psi\to\gamma\eta_c$.

However, the (semi-)quantitative results that we obtained within the WSS model suggest that the experimental identification of the pseudoscalar
glueball may be made difficult by a very large decay width if its mass is around 2.6 GeV as indicated by (quenched) lattice results.
In future work we intend to further explore these predictions in the context of radiative $J/\psi$ decays and central exclusive production
in the Regge regime.

\acknowledgments

We thank Francesco Bigazzi for useful discussions.
J.L.\ has been supported by the doctoral programme DKPI 
of the Austrian Science Fund FWF, project no.\ W1252-N27.


\bibliographystyle{JHEP} 

\bibliography{glueballdecay}


%
%
%
%
%
%
%
%
\end{document}